\documentclass[preprint,showpacs,preprintnumbers,amsmath,amssymb]{revtex4}
\usepackage{graphicx}
\usepackage{dcolumn}
\usepackage{bm}
\usepackage[usenames]{color}
\newcommand{\be}{\begin{equation}}
\newcommand{\en}{\end{equation}}
\newcommand{\bea}{\begin{eqnarray}}
\newcommand{\ena}{\end{eqnarray}}

\usepackage{perpage} 
\MakePerPage{footnote} 

\begin{document}


\renewcommand\a{\alpha}
\renewcommand\b{\beta}
\renewcommand\c{\chi}
\renewcommand\d{\delta}
\newcommand\D{\Delta}
\newcommand\eps{\epsilon}
\newcommand\vareps{\varepsilon}
\newcommand\g{\gamma}
\newcommand\G{\Gamma}
\newcommand\grad{\nabla}
\newcommand\h{\frac{1}{2}}
\renewcommand\k{\kappa}
\renewcommand\l{\lambda}
\renewcommand\L{\Lambda}
\newcommand\m{\mu}
\newcommand\n{\nu}
\newcommand\om{\omega}
\renewcommand\O{\Omega}
\newcommand\p{\phi}
\newcommand\vp{\varphi}
\renewcommand\P{\Phi}
\newcommand\pa{\partial}
\newcommand\tpa{{\tilde \partial}}
\newcommand\bpa{{\bar \partial}}
\newcommand\pr{\prime}
\newcommand\ra{\rightarrow}
\newcommand\lra{\longrightarrow}
\renewcommand\r{\rho}
\newcommand\s{\sigma}
\renewcommand\S{\Sigma}
\renewcommand\t{\tau}
\renewcommand\th{\theta}
\newcommand\bth{{\bar \theta}}
\newcommand\Th{\Theta}
\newcommand\z{\zeta}
\newcommand\ti{\tilde}
\newcommand\wti{\widetilde}

\newcommand\rf[1]{(\ref{eq:#1})}
\newcommand\lab[1]{\label{eq:#1}}
\newcommand\nonu{\nonumber}
\newcommand\br{\begin{eqnarray}}
\newcommand\er{\end{eqnarray}}
\newcommand\ee{\end{equation}}
\newcommand\eq{\!\!\!\! &=& \!\!\!\! }
\newcommand\foot[1]{\footnotemark\footnotetext{#1}}
\newcommand\lb{\lbrack}
\newcommand\rb{\rbrack}
\newcommand\llangle{\left\langle}
\newcommand\rrangle{\right\rangle}
\newcommand\blangle{\Bigl\langle}
\newcommand\brangle{\Bigr\rangle}
\newcommand\llb{\left\lbrack}
\newcommand\rrb{\right\rbrack}
\newcommand\Blb{\Bigl\lbrack}
\newcommand\Brb{\Bigr\rbrack}
\newcommand\lcurl{\left\{}
\newcommand\rcurl{\right\}}
\renewcommand\({\left(}
\renewcommand\){\right)}
\renewcommand\v{\vert}                     
\newcommand\bv{\bigm\vert}               
\newcommand\Bgv{\;\Bigg\vert}            
\newcommand\bgv{\bigg\vert}              
\newcommand\lskip{\vskip\baselineskip\vskip-\parskip\noindent}
\newcommand\mskp{\par\vskip 0.3cm \par\noindent}
\newcommand\sskp{\par\vskip 0.15cm \par\noindent}
\newcommand\bc{\begin{center}}
\newcommand\ec{\end{center}}
\newcommand\Lbf[1]{{\Large \textbf{{#1}}}}
\newcommand\lbf[1]{{\large \textbf{{#1}}}}

\title{ Instant preheating  in a scale invariant two measures theory}

\author{Eduardo I. Guendelman}

\email{guendel@bgu.ac.il} 
\affiliation{ Physics Department, Ben Gurion University, Beer Sheva, Israel.}

\author{Ram\'on Herrera}

\email{ramon.herrera@pucv.cl} \affiliation{ Instituto de
F\'{\i}sica, Pontificia Universidad Cat\'{o}lica de
Valpara\'{\i}so, Avenida Brasil 2950, Casilla 4059,
Valpara\'{\i}so, Chile.}

\author{Pedro Labra$\tilde{n}$a}
\email{plabrana@ubiobio.cl} \affiliation{ Departamento de F\'{\i}sica, Universidad del B\'{\i}o-B\'{\i}o, 
Casilla 5-C, 
Concepci\'on, Chile.}

\date{\today}
\begin{abstract}
The instant preheating mechanism in the framework of a scale 
invariant two measures theory  is studied. We  introduce this 
mechanism into a non oscillating   inflationary model as another possible solution to the
reheating of the universe  in this   theory.  In this framework, we consider that the model  
includes 
two scalar matter fields, the first a dilaton field, that transforms under scale 
transformations and it will be considered also as the field that drives 
inflation 
and the second, a scalar field which will interact with the inflaton
 through an effective potential.
By assuming this interaction term, 
we obtain a scenario of instant radiation  or decay of particles according to the domain 
the effective mass of the 
field that interacts with the inflaton.
Also,  we consider a scale invariant 
Yukawa interaction and then after performing the transition 
to the physical Einstein frame we obtain an expression for the decay rate from our 
scalar field going into two fermions.
Besides,  
from specific  decay rates, different constraints and  bounds 
for the coupling parameters associated with our model are found.

\end{abstract}

\maketitle

\section{Introduction}

It is well known that the
inflationary universe models  have solved some problems present in  the  hot big bang model, such 
as the horizon, flatness, monopole problem  etc.\cite{1,2,3}. However, the biggest feature of the inflationary
stage is that it provides a causal interpretation to explicate  the observed anisotropy of the
cosmic microwave background (CMB) radiation\cite{CMB} and moreover this framework gives account of the
distribution of large scale structures \cite{LSS}.

In order to study the inflationary stage, the scalar field or inflaton plays a fundamental role in
the evolution of the early universe. In this context, the inflationary epoch  can be described  as a 
regime with a rapid  accelerated expansion occurred during  the early  universe produced   by the 
inflaton\cite{1,2}. 

To describe the inflationary epoch, we have  
different gravitation theories and models  that give account of the evolution of the early universe. 
In particular we
can distinguish the scale invariant  two measures theory\cite{GG1,GG2,GG3} that produces    
an   accelerated expansion of the  universe by means of the evolution of a single scalar field or inflaton 
field with an effective potential \cite{Guendelman:2014bva,G2,G3}.

In relation to the  two measures theories models, these  utilize a non Riemannian measure of integration in 
the frame of the action. In particular 
in the  situation of a scale invariant theory, the scale invariance was spontaneously broken from  the 
equations of motion related with the degrees of freedom on  the non Riemannian measure of integration in 
the framework of the action. In this sense, we can mention that 
the degrees of freedom that determine a non Riemannian measure of integration in four dimensions can be 
represented by  scalar fields\cite{Guendelman:2014bva,G2}. 
In this sense, utilizing   the measure of integration and also in the frame of  the action  different models with 
several scalar fields in four dimensions have been studied in the literature\cite{St,Dc}. The application of this  
scale invariant  two measures theory to an emergent universe scenario was developed in ref.\cite{Dc}. 
It corresponds to a non singular cosmological type of stage previous to inflation (emergent scenario) in 
which the universe begins as a static  universe to later connect with the 
 inflationary epoch\cite{Em}. In order to use the two measures theory to 
 describe the dark energy in the present universe, in ref.\cite{Em} was  considered that the 
two measures of integration leading to two independent integration constants and 
these  constants break scale invariance, and characterize  the strength of the
 dark energy density. Additionally,  in ref. \cite{Guendelman:2015liz}
the curvaton reheating 
mechanism in a 
scale 
invariant two measures theory defined in terms of two independent non-Riemannian 
volume forms was studied.  In this context, 
the model has two scalar matter fields, a dilaton and it transforms under 
scale transformations and it corresponds to the inflaton  field  of the inflationary model while the other scalar field 
 does not transform under scale transformations  played the role of a curvaton 
 field\cite{Guendelman:2015liz}.   The introduction of the curvaton field in this 
 scenario occurs due to the 
problematic of  connecting   the 
inflationary epoch in the framework of the scale invariant two measures theory, 
with the reheating  of the universe and  its subsequent connection with the radiation 
era \cite{Cu,Cu1}.

In relation to the reheating of the universe, we have  that  at the end of inflationary epoch
 the energy density of the universe can be interpreted as 
 a combination of kinetic and potential energies of the inflaton to late dominate the kinetic energy  
\cite{Re}.
In the process of  reheating of the universe,  the matter and radiation of the universe 
are produced   generally  
through  the decay of the scalar field or another field    (decay parameter), 
while the temperature increases in 
many orders of magnitude and then 
 the universe connects with the radiation regime of the standard big-bang 
 model\cite{Re1,Feng}.

In order to study the reheating of the universe 
the scenario of oscillations of the inflaton field (at the minimum of the potential) is an important  part 
for the standard mechanism of reheating. 
Nevertheless, 
it 
  is possible 
to find  some inflationary 
models where the effective  potential associated to the inflaton 
 does not have a minimum and then the scalar 
field does not oscillate and then the standard mechanism of reheating does not 
work\cite{Ko}. Thus, these kinds of models with these  effective potentials 
are known in the literature as  non-oscillating models, or 
simply NO models\cite{Fe}.
Interesting examples of these are the Quintessential inflation models which connect an early inflation 
with a late slowly accelerated phase, as the models considered by Peebles and Vilenkin\cite{Pee} as well as
\cite{GG3,Guendelman:2014bva,G2,Dc,Jau,Be}. 

Originally, in order to solve this problematic for these NO models  was
the introduction of a mechanism that incorporates  the gravitational particle 
production\cite{Fo}. Nevertheless, this mechanism of reheating of the universe becomes 
inefficient and it presents several problems associated with 
 the observational data, see ref.\cite{Ur}.

The introduction of the curvaton field 
as 
other mechanism of reheating after inflation in these NO models was considered in refs. 
\cite{Cu,Cu1,Fe,Feng}. In this context, 
  the decay rate of the scalar  curvaton 
field into conventional matter gives account
 a mechanism of reheating of the universe.  In this sense, introducing an 
 effective potential associated to the
 curvaton field is possible to reheat the universe\cite{Feng,Campuzano:2006eu}.
 This model of reheating   does not need to
  introduce an interaction between the scalar field 
that drives   inflation and other scalar field\cite{Feng,Cu1}.

Another mechanism of reheating known in the literature  is called 
instant preheating\cite{Fe1}. In this scenario, after of the inflationary regime  
the inflaton field moves  quickly  producing particles 
which can be bosons and/or fermions.  This mechanism corresponds to 
a non-perturbative process and it happens  almost instantly\cite{Fe1} 
 and also this scenario does not need  oscillations or parametric resonance 
 of the inflaton field. 
In this sense, because 
the 􏰟production 
of 􏰶  􏰵 particles􏰎 can happen  immediately after the end of inflationary regime, within less 
than one oscillation of the  field that drives inflation, the reheating of the universe can 
occurs   efficiently. 
 In order to 
 study the instant reheating is indispensable to consider the interaction 
 between the scalar field that drives inflation and another scalar field $\sigma$. 
Depending of the interaction between the inflaton field and the  field  $\sigma$
(via an effective  potential) the effective masses of the particles $\sigma$ can 
be small or large at the moment  when the particles are produced for later 
increase or decrease when the inflaton field moves to large values.  In this mechanism the 
production of particles $\sigma$ begins nearly instantaneously assuming the 
nonadiabatically condition given by the ratio between the evolution of the 
effective masses of the particles-$\sigma$ and the square of these\cite{Fe1}.
For a 
review of reheating see  refs.\cite{Fe1,Shtanov:1994ce,Reh2} and for 
instant preheating, see \cite{Inst,Jau}.

The goal of this investigation is to analyze the instant preheating in a scale 
invariant two independent non Riemannian volume-forms.
In this sense, we investigate how the interaction term between the inflaton field that drives inflation
and other scalar field (from the effective 
potential) in this theory 
modifies the results on the produced particles in this scenario and preheating of the universe. 
In this form, we will analyze the instant preheating in our model and in particular the 
energy density of produced particles and the 
decay rates in order to in account of the temperature and constraints on the 
parameters  given by the observations.

For the application of the developed formalism, we will analyze some examples 
assuming two  decay parameters. From these decay rates, we will study the different 
conditions of time, in order to obtain the bounds on the coupling parameters associated to 
these decay rates.

The outline of the paper goes as follow: in Sect. II we give a brief 
description of two independent non-Riemannian volume-forms. In Sect. III 
the instant preheating scenario is 
analyzed. The Sect. IV describes the instant radiation in which the energy 
density of the field-$\sigma$ decays as radiation. The Sect. V explains the 
radiation from the decay. In Sect. VI we obtain the decay rate for the particles $\sigma$ 
going into two fermions. The Sect. VII analyzes the decay rate and constraints 
on the parameters of our model,   and in Sect. VIII includes our conclusions.



\section{Two independent non-Riemannian volume-forms}

In this section, we discuss a brief description of the   two independent non-Riemannian volume-forms. 
We follow the general structure of the references  \cite{Guendelman:2014bva,G2}, but now we will enrich 
the field content
of the theory with a new field $\sigma$ which will not transform under scale transformations, so we write,
\label{TMMT}
\begin{equation}
S = \int d^4 x\,\Phi_1 (A) [ R + L^{(1)}]  +
\int d^4 x\,\Phi_2 (B)\left[ L^{(2)} + \epsilon R^2 +
\frac{\Phi (H)}{\sqrt{-g}}\right] \; ,
\lab{TMMT}
\end{equation}
where $\Phi_1(A)$ and $\Phi_2(B)$ are two independent non-Riemanniam volume-forms and defined as
\be \P_1 (A) = \frac{1}{3!}\vareps^{\m\n\k\l} \pa_\m
A_{\n\k\l}, \quad \mbox{and}\quad \P_2 (B) = \frac{1}{3!}\vareps^{\m\n\k\l}
\pa_\m B_{\n\k\l} \; , \lab{Phi-1-2} \ee 
respectively. 

The quantities 
$L^{(1,2)}$ correspond to two different
Lagrangians of two scalar fields, the dilaton $\vp$, which will
play the role of an inflaton and an additional scalar field  $\sigma$.
In this form, the Lagrangians can be written as
$$
L^{(1)} = -\h g^{\m\n} \pa_\m \vp
\pa_\n \vp -\h g^{\m\n} \pa_\m \sigma \pa_\n \sigma -\frac{\mu^2
\sigma^2}{2} \exp \{-\a\vp\} -  V(\vp), 
$$
where $V(\vp) = f_1 \exp
\{-\a\vp\}$  and the Lagrangian 
$$
L^{(2)} = -\frac{b}{2} e^{-\a\vp} g^{\m\n} \pa_\m \vp \pa_\n \vp 
+ U(\vp),
$$
in which  $U(\vp) = f_2 \exp \{-2\a\vp\} $.
Here the quantities $\a, f_1, f_2$ are dimension full positive
parameters and the parameter $b$ is a dimensionless one. Also, the quantity
$\P (H)$
denotes the dual field strength of a third auxiliary 3-index
antisymmetric tensor gauge field defined as \be \P (H) =
\frac{1}{3!}\vareps^{\m\n\k\l} \pa_\m H_{\n\k\l} \; . \lab{Phi-H}
\ee 

We mention that the scalar potentials have been chosen 
such that the  action given by 
eq.\rf{TMMT} is invariant under global Weyl-scale transformations with which
$$
g_{\m\n} \to \l g_{\m\n} \;\; ,\;\; \G^\m_{\n\l} \to \G^\m_{\n\l} \;\; ,\;\;
\vp \to \vp + \frac{1}{\a}\ln \l \;\;, \sigma  \to \sigma,
$$
$$
A_{\m\n\k} \to \l A_{\m\n\k} \;\; ,\;\; B_{\m\n\k} \to \l^2 B_{\m\n\k}
\;\; ,\;\; H_{\m\n\k} \to H_{\m\n\k} \; .
$$
Analogously, from the invariant under we have multiplied by an  exponential factor
the scalar kinetic term in $L^{(2)}$ and also by the scalar curvature $R$ and $R^2$ couple to the two
different modified measures.
The equations of motions of the measure fields lead to several simple relations. First, the variation of 
the tensor field $H$ implies that the ratio between the measure $\Phi_2$ and $\sqrt{-g}$ is a constant:
\begin{equation}
\frac{\Phi_2}{\sqrt{-g}} = \chi_2 = \mbox{constant}.
\end{equation}
\label{measures-ratio}
Likewise, the variation with respect to $\Phi_1$ and  $\Phi_2$ leads to the the Lagrangians coupling 
to  $\Phi_1$ and  $\Phi_2$
being constants that we may call $M_1$ and $M_2$ :
\begin{equation}
 R + L^{(1)} =- M_1 = \mbox{constant},\label{breaking scale 1}
\end{equation}
and 
\begin{equation}
  L^{(2)} + \epsilon R^2 +
\frac{\Phi (H)}{\sqrt{-g}} =- M_2 = \mbox{constant},\label{breaking scale 2}
\end{equation}
while equation (\ref{measures-ratio}) does not break scale invariance, since the two measures $\Phi_2$ 
and $\sqrt{-g}$ transform identically under scale transformations. The same cannot be said however 
concerning (\ref{breaking scale 1}) and 
(\ref{breaking scale 2}) while the left hand side in these equations transforms, the right hand side 
($M_1$ and $M_2$ )are  constants and does not transform. We get then spontaneous breaking of scale 
invariance.

We proceed in the so called first order formalism, where the connection is at the action level independent
 of the metric, in this case we can vary with respect to the metric and the consistency with the equations 
 (\ref{breaking scale 1}) and (\ref{breaking scale 2}) allows us to solve for  $\frac{\Phi_1}{\sqrt{-g}} = \chi_1$,
which is given by,
\begin{equation}
\frac{1}{\chi_1} =\frac{(V+\frac{\mu^2\sigma^2}{2} e^{-\alpha\varphi}-M_1)}{2\chi_2(U+M_2)}.\label{chi1}
\end{equation}
Here we have considered  the case in which  $\epsilon=0$ and $b=0$, respectively.

Defining Einstein frame by a conformal transformation,
we obtain an effective action, which in the case of $\epsilon=0$ and $b=0$ is governed by a canonical 
minimally coupled scalar field with the following effective Lagrangian given by 
\begin{equation}
L_{eff}=-\frac{1}{2}\bar{g}^{\mu\nu}\partial\varphi_\mu\partial\varphi_\nu-
\frac{1}{2}\bar{g}^{\mu\nu}\partial\sigma_\mu\partial\sigma_\nu-U(\varphi,\sigma),
\end{equation}
where the Weyl-rescaled metric $\bar{g}_{\mu\nu}$ is defined as 
\begin{equation}
\bar{g}_{\mu\nu}=\chi_1\;g_{\mu\nu},
\end{equation}
and the effective potential is given by 
\begin{equation}
U_{\rm eff} (\varphi, \sigma)=
\frac{(V+ \frac{\mu^2 \sigma^2}{2}\; e^{-\alpha\varphi} - M_1)^2}{4\chi_2 [ U + M_2]}= 
\frac{(f_1 e^{-\alpha\varphi}+ \frac{\mu^2 \sigma^2}{2}\; e^ {-\alpha\varphi} -M_1)^2}{4\chi_2\,[
f_2\; e^{-2\alpha\varphi} + M_2 ]}. \label{UEFF}
\end{equation}

\section{Instant preheating}
In order to explain the instant preheating scenario in our model we will consider that the effective potential
given by Eq.(\ref{UEFF}) presents the interaction term given by 
\begin{equation}
U_{\rm eff} (\varphi, \sigma)\approx \frac{\mu^2\beta^2\,\sigma^2}{2}\:e^{-2\alpha\varphi}=
\frac{m_1^2\;\sigma^2}{2}\:e^{-2\alpha\varphi},\label{pot1}
\end{equation}
where the constant $\beta$ is defined as
$\beta^2=\frac{f_1}{2\chi_2M_2}$ and $m_1=\mu\beta$. Here we have assumed  that 
$M_2\gg f_2 e^{-2\alpha\varphi}$ and $f_1e^{-\alpha\varphi}\gg M_1$, since during the 
inflationary scenario  we have used the values $M_1\sim10^{-60}$, $f_1\simeq f_2\sim 10^{-8}$ and 
$M_2\sim 1$, from observational data, see ref.\cite{Guendelman:2014bva}.
Thus, the effective mass of the scalar field $\sigma$ becomes  
\begin{equation}
m_{\sigma}=\mu\beta\;e^{-\alpha\varphi}=m_1\,e^{-\alpha\varphi},
\end{equation}
since the effective mass of $\sigma$ is defined as $m_\sigma^2=\partial^2
U_{\rm eff}(\varphi,\sigma)/\partial\sigma^2$.

Following refs.\cite{LNO,2LNO} we will consider that 
 the production of particles $\sigma$ starts to change nonadiabatically under the condition 
$|\dot{m_\sigma}|\geq m_\sigma^2$ with which the scalar field $\varphi$ can be written as 
\begin{equation}
\varphi\sim -\frac{1}{\alpha}\;\ln\left(\frac{\alpha |\dot{\varphi}_{0}|}{\mu\beta}\right),
\end{equation}
where $\dot{\varphi}_{0}$ denotes the value of the velocity of the scalar field when this field  rolls on 
the asymptotically   flat potential after of the inflationary epoch.
During this stage the mechanism of particle production starts nearly instantaneously in the time  interval 
given by 
\begin{equation}
\triangle\,t\sim\,\frac{|\varphi|}{|\dot{\varphi_0}|}\sim \frac{1}{\alpha|\dot{\varphi_0}|}
\;\big|\ln\left(\frac{\alpha |\dot{\varphi}_{0}|}{\mu\beta}\right)\big|>0.\label{TT}
\end{equation}
Also, we mention that during this time all effects associated to the expansion of the universe can 
be ignored in the process of particle production.

Now, in order to determine the velocity of the scalar field $\dot{\varphi}_0$, we can consider the break 
down approximation in which 
\begin{equation}
\ddot{\varphi}\simeq -\frac{\partial V(\varphi)}{\partial \varphi}=\frac{\alpha\,f_1^2}{2\chi_2M_2}
e^ {-2\alpha\varphi},\label{BDE}
\end{equation}
where we have considered that the potential  $V(\varphi)=\frac{f_1^2}{4\chi_2M_2}
e^ {-2\alpha\varphi}$ from  eq.(\ref{UEFF}), see ref.\cite{Guendelman:2014bva}.

Under this approximation, we  find that the solution of the eq.(\ref{BDE}) for the scalar field $\varphi(t)$ results
\begin{equation}
\varphi(t)=\frac{1}{\alpha}\,\ln\left[ \frac{e^{-\alpha\sqrt{C_1}(t+C_2)}}{2}\left(k_1^2+
\frac{e^{2\alpha\sqrt{C_1}(t+C_2)}}{\alpha C_1}\right)   \right],\label{EQ1}
\end{equation}
where $C_1$ and $C_2$ are two integration constants and $k_1$ is defined as $k_1^2=
\frac{\alpha f_1^2}{2\chi_2M_2}$. 

From this solution we can find that the velocity of the scalar field $\dot{\varphi}$ is  given by 
\begin{equation}
\dot{\varphi}(t)=\frac{\sqrt{C_1}\,(e^{2\alpha\sqrt{C_1}(t+C_2)}-\alpha C_1k_1^2)}
{e^{2\alpha\sqrt{C_1}(t+C_2)}+\alpha C_1k_1^2},\label{eq2}
\end{equation}
which for  $\alpha$ big the above expression big quickly approaches the asymptotic value
$\sqrt{C_1}$ i.e.,  $\dot{\varphi_0}\sim\sqrt{C_1} $.

Thus, in order to obtain the value of the  $\dot{\varphi_0}$, we can  consider that the initial 
conditions for the scalar field and its velocity can be fixed at the end of inflationary epoch. 
In this way, we assume  the slow roll approximation in which at the end of inflation we have 
$\varphi_{end}=-\alpha^{-1}\ln(2\alpha M_1/f_1)$ and $\dot{\varphi}_{end}=
\frac{2M_1\alpha^2}{\sqrt{3\chi_2M_2}}$, see ref.\cite{Guendelman:2014bva}. 
From these initial conditions  and considering the eqs.(\ref{EQ1}) and (\ref{eq2}), 
we find that the asymptotic velocity of the scalar field becomes
\begin{equation}
\dot{\varphi}_0\simeq\,\sqrt{C_1}\,\simeq\dot{\varphi}_{end}\;\,\sqrt{1+\frac{3}{2\alpha^2}}.\label{eq18}
\end{equation}
Here we note that for large-$\alpha$ the velocity 
$\dot{\varphi_0}\simeq\dot{\varphi}_{end}$. Thus, the time interval given by eq.(\ref{TT}) 
can be approximated to $\triangle\,t\sim\, (\alpha|\dot{\varphi}_{end}|)^{-1}\;
\ln\left(\frac{\alpha |\dot{\varphi}_{end}|}{\mu\beta}\right)$.

On the other hand, the occupation number $n_k$ of the particles $\sigma$ with momentum 
$k$ in the time interval $\triangle\,t$ is defined as \cite{LNO,2LNO,Sta}
\begin{equation}
n_k=\exp[-\pi\,(k\,\triangle\,t)^2],\label{AA}
\end{equation}
and then considering eq.(\ref{TT}) we obtain  that the occupation number can be written as
\begin{equation}
n_k=\exp\left[-\frac{\pi\,k^2}{\alpha^2\,\dot{\varphi_0}^2}\,\left(\ln\left[\frac{\alpha|\dot{\varphi_0}|}
{\mu\beta}\right]\right)^2\right].\label{MS}
\end{equation}
In fact, we can  assume  that the definition of the occupation number given by eq.(\ref{AA}) still is 
valid for massive particles of the scalar field $\sigma$ of effective mass $m_\sigma$ under replacement of 
 the momentum $k^2$ by $k^2+m_\sigma^2$ \cite{2LNO}. Thus, eq.(\ref{AA}) can be modified as  
 $n_k=\exp[-\pi\,(k^2+m_\sigma^2)\,\triangle\,t^2]$ with  which the occupation number becomes
\begin{equation}
n_k=\exp\left[-\frac{\pi\,(k^2+m_\sigma^2)}{\alpha^2\,\dot{\varphi_0}^2}\,
\left(\ln\left[\frac{\alpha|\dot{\varphi_0}|}{\mu\beta}\right]\right)^2\right].
\end{equation}
Now, this quantity can be integrated to establish the density of $\sigma$ particles denotes by 
$n_\sigma$ and  defined as $n_\sigma=\frac{1}{2\pi^2}\int_0^\infty\,d k\,k^2\,n_k$. 
In this way, we find that the density of $\sigma$ particles $n_\sigma$ results
\begin{equation}
n_\sigma=\frac{1}{8\pi^3}\,\left[\frac{\alpha\,|\dot{\varphi_0}|}{\big|\ln\left(\frac{\alpha\,|\dot{\varphi_0}|}
{\mu\beta}\right)\big|}\right]^{3}\,\exp\left[-\frac{\pi\,m_\sigma^2}{\alpha^2\,\dot{\varphi_0}^2}\,
\left(\ln\left[\frac{\alpha|\dot{\varphi_0}|}{\mu\beta}\right]\right)^2\right].
\end{equation}

We note that naturally  in our model the number of produced particles is not exponentially 
suppressed, since the mass of the scalar field $\sigma$ decreases  for large-$\varphi$ 
($m_\sigma\propto e^{-\alpha \varphi}$).
Thus, the number density of particles during their creation results $\frac{1}{8\pi^3}\,
\left[\frac{\alpha\,|\dot{\varphi_0}|}{\big|\ln\left(\frac{\alpha\,|\dot{\varphi_0}|}{\mu\beta}\right)\big|}\right]^{3}$, 
however it decreases as $a^{-3}(t)$ with which   the number of produced particles in terms of the time can 
be written as
\begin{equation}
n_\sigma=\frac{1}{8\pi^3}\,\left[\frac{\alpha\,|\dot{\varphi_0}|a_0}{a(t)\,\big|\ln\left(\frac{\alpha\,|\dot{\varphi_0}|}
{\mu\beta}\right)\big|}\right]^{3}.
\end{equation}
Here we have used that at the moment of particle production  the scale factor is given by  $a_0$.

Additionally, we have that the energy density of produced particles $\rho_\sigma$ is defined as
\cite{L1,Dolgov:2014jaa}
\begin{equation}
\rho_\sigma=\frac{1}{(2\pi\,a)^3}\int_0^\infty\,\,n_k\,\,\sqrt{\frac{k^2}{a^2}+m_\sigma^2}\,\,
\;(4\pi k^2) d k.\label{r1}
\end{equation}
Here, we can  note that interestingly  there are two limit cases 
given by  $m_\sigma\gg k/a$ and $m_\sigma\ll k/a$, because the effective  mass of the $\sigma$-field  
 $m_\sigma$ depends of the time. Thus, initially after of the inflationary stage, we can consider that the 
dominant term 
becomes the mass $m_\sigma$ over the physical momentum $k/a$. Later, product of the decrease in the 
time of the mass $m_\sigma$, 
the dominant term corresponds to the momentum i.e.,  $k/a\gg m_\sigma$.
In the following, we will analyze these two limits separately.

\section{Instant radiation}

 In this section we will study the process in which the energy density of produced particles of the field 
 $\sigma$ decays as radiation. We call  this process as 
 instant radiation and it  occurs for large-time  when the mass of the $\sigma-$field decreases and then the
 effective mass tends to zero  with which 
$m_\sigma\ll k/a$. In this situation we find that energy density of the $\sigma-$field from Eq.(\ref{r1}) 
becomes
\begin{equation}
\rho_\sigma=B a^{-4},\label{rad}
\end{equation}
where the constant $B$ is 
defined as
$$
B
=\left[\frac{\alpha\,\dot{\varphi_0}}{\left
(2^{1/2}\pi\ln
\left[\frac{\alpha\,|\dot{\varphi_0}|}{\mu\beta}\right]\right)}\right
]^4.
$$

In this context, the equation of motion for the inflation field $\varphi$
including backreaction of produced $\sigma$ particles on the field $\varphi$ can be written as 
\begin{equation}
\ddot{\varphi}+3H\dot{\varphi}=\alpha\mu^2\beta^2\,e^{-2\alpha\varphi}\langle \sigma^2\rangle,\label{dd}
\end{equation}
where the expectation value $\langle \sigma^2\rangle$ is defined as \cite{Vilenkin:1982wt}
\begin{equation}
\langle \sigma^2\rangle\approx\frac{1}{2\pi^2}\int\frac{n_k\, k^2\, d k}{\sqrt{(k/a)^2+m_\sigma^2}}\label{27}.
\end{equation}
Thus, for the case of the instant radiation ($m_\sigma\ll k/a$) we find that $\langle \sigma^2\rangle$ becomes

\begin{equation}
\langle \sigma^2\rangle\approx\frac{1}{2\pi^2}\int n_k\, k\, d k=\,\frac{B_1}{a^{2}(t)},\label{b27}
\end{equation}
where the constant $B_1$ is given by $B_1=\sqrt{B}/(2\pi)$.

In this way, the  equation of motion for the inflaton field  in the situation in which the effective mass  
$ m_\sigma \ll k/a$ including the backreaction term becomes
\begin{equation}
\ddot{\varphi}+3H\dot{\varphi}=\alpha\,\mu^2\,\beta^2\,B_1\,
\frac{e^{-2\alpha\varphi(t)}}{a^2(t)}.\label{b28}
\end{equation}
 We observe that the backreaction effect decreases very quickly due to exponential decay product of 
 evolution of $\varphi(t)$  that appears in the right hand side of this equation. Thus, the backreaction of 
 produced $\sigma$ particles on   $\varphi$ disappears naturally from the effective potential given by eq.(\ref{pot1}).
Also, from the condition  $m_\sigma\ll k/a$ and considering that the scale factor $a(t)\propto t^{1/3}$ 
together with  neglecting  the backreation of eq.(\ref{b28}), we find that the constraint for the $\alpha$ 
parameter becomes $\alpha>(2\sqrt{3})^{-1}$, if we want the condition $m_\sigma\ll k/a$ to be maintained  
during the time. 

Additionally, 
we note that in the scenario of instant radiation, if nothing else happens, meaning non-decay of the 
$\sigma$ particles, and  since the mass of these particles approaches to zero, then we obtain that 
these particles will asymptotically behave as radiation, as can be seen from  eq.(\ref{rad}). However, we 
mention that this spectrum is not thermal becomes the distribution in the occupation number is not 
Boltzmann distribution (see eq.(\ref{AA})), since the spectrum is 
not thermal in order to obtain a real 
thermal spectrum a thermalization process is required. 
The thermalization should bring all particle species\cite{Ed,Ed1}.

\section{Radiation from decay}

In this section, we 
can analyze the  case  where the mass $m_\sigma\gg k/a$,  for the dominant range of integration of the 
momentum. In this limit we have 
\begin{equation}
\rho_\sigma=m_\sigma\,n_\sigma=\frac{\mu\beta}{8\pi^3}\,\left[\frac{\alpha\,|\dot{\varphi_0}|a_0}
{\big|\ln\left(\frac{\alpha\,|\dot{\varphi_0}|}{\mu\beta}\right)\big|}\right]^{3}\,\frac{e^{-\alpha\,\varphi(t)}}{a^{3}(t)}
\propto\frac{ e^{-\alpha\,\varphi(t)}}{a^{3}(t)},\label{25}
\end{equation}
here, we have called to this stage as radiation from decay.

On the other hand, the equation of motion for the inflation field $\varphi$ after of the particles production  
can be written as 
\begin{equation}
\ddot{\varphi}+3H\dot{\varphi}=\alpha\mu^2\beta^2\,e^{-2\alpha\varphi}\langle \sigma^2\rangle,\label{dd}
\end{equation}
where the expectation value $\langle \sigma^2\rangle$ from eq.(\ref{27}) and assuming $m_\sigma\gg k/a$ 
can be written as
\begin{equation}
\langle \sigma^2\rangle\approx\frac{1}{2\pi^2}\int\frac{n_k\, k^2\, d k}{\sqrt{(k/a)^2+m_\sigma^2}}\approx
\frac{n_\sigma}{m_\sigma}\approx\,A\,\,\frac{e^{\alpha\varphi(t)}}{a^{3}(t)},\label{27b}
\end{equation}
in which the constant $A$ is defined as
$$
A=\frac{1}{8\mu\beta\pi^3}\,\left[\frac{\alpha\,|\dot{\varphi_0}|a_0}{\big|\ln\left(\frac{\alpha\,|\dot{\varphi_0}|}
{\mu\beta}\right)\big|}\right]^{3}.
$$

In this form, using eq.(\ref{27b}) we find that eq.(\ref{dd}) can be rewritten as 
\begin{equation}
\ddot{\varphi}+3H\dot{\varphi}=\alpha\, m_\sigma\,\,n_\sigma=\frac{\alpha\mu\beta}{8\pi^3}\,
\left[\frac{\alpha\,|\dot{\varphi_0}|a_0}{\,\big|\ln\left(\frac{\alpha\,|
\dot{\varphi_0}|}{\mu\beta}\right)\big|}\right]^{3}\frac{e^{-\alpha\varphi(t)}}{a^3(t)}.\label{28}
\end{equation}

In order to analyze  the behavior of the field  $\varphi(t)$ from eq.(\ref{28}), we will consider as a first 
approximation neglects backreaction of produced particles. In this context, we can
consider that the energy density of the $\sigma-$particles ($\rho_\sigma=m_\sigma\,n_\sigma$) becomes
subdominant and then the right hand side of eq.(\ref{28}) can be negligible. In fact, from eq.(\ref{28}) 
(or analogously of eq.(\ref{dd})),  we note that naturally our effective potential gives rise to a force which 
produces that  
the inflaton field continues its movement to infinity.  

Under this approximation  the scale factor is given by $a(t)\sim t^{1/3}$
 and then the Hubble parameter becomes $H=(3t)^{-1}$. 
 Thus, if one neglects backreaction, the solution for the scalar field as a function of the time can be written as  
\begin{equation}
\varphi(t)=\frac{2}{\sqrt{3}}\,\ln\left(\frac{t}{t_0}\right),\label{EQ26}
\end{equation}
where the constant $t_0$ is defined as $t_0=(3H_0)^{-1}$
and it corresponds to the initial time   during the phase transition time between inflation and kination regime. 

In this way, replacing eq.(\ref{EQ26}) in the equation for the energy density of 
produced particles given by eq.(\ref{25}) we have
\begin{equation}
\rho_\sigma=\alpha_1\left(\frac{t}{t_0}\right)^{-2\alpha/\sqrt{3}}\,\left(\frac{a_0}{a}\right)^3,
\end{equation}
where the constant $\alpha_1$ is defined as
$$
\alpha_1=\frac{\mu\beta}{8\pi^3}\,\left[\frac{\alpha\,|\dot{\varphi_0}|}
{\big|\ln\left(\frac{\alpha\,|\dot{\varphi_0}|}{\mu\beta}\right)\big|}\right]^{3}.
$$

Additionally,  during  the kination regime the energy density of the background 
decreases as $\rho(t)\sim\dot{\varphi}^2\sim a^{-6}$ or 
$\rho(t)=6H_0^2\,(a_0/a)^6$ and we can consider that 
both densities achieve  equilibrium i.e.,  $\rho\sim \rho_\sigma$. In this sense, if 
the  densities $\rho$ and $\rho_\sigma$ are of the same order, we can assume  that this situation occurs 
at the equilibrium time $t_{eq}$ given by   
\begin{equation}
t_{eq}=\left[\frac{2}{3\alpha_1}\,t_0^{\delta_1}\right]^{1/\delta_2},\label{eq1}
\end{equation}
where the constants $\delta_1$ and $\delta_2$ are defined as
$$
\delta_1=-(1+2\alpha/\sqrt{3}),\,\;\;\;\;\mbox{and}\,\,\,\,\delta_2=(1-2\alpha/\sqrt{3}),
$$
respectively. In this way, the value of the scalar field at the time $t_{eq}$ becomes
\begin{equation}
\varphi(t=t_{eq})=\varphi_{eq}=\frac{2}{\sqrt{3}\;\delta_2}\left[\ln\left(\frac{2}{3\alpha_1}\right)
+(\delta_1-\delta_2)\ln\,t_0)\right].\label{t1}
\end{equation}
We note that in particular for values of $\alpha\gg 1$, we have
\begin{equation}
\varphi_{eq}\simeq-\frac{1}{\alpha}\,\ln\left[\frac{2}{3\alpha_1\,t_0^2}\right]=-\frac{1}{\alpha}\,
\ln\left[\frac{\dot{\varphi_0}^2}{2\alpha_1}\right].\label{P1}
\end{equation}
Here, we have used that $t_0=2/(\sqrt{3}\,\dot{\varphi_0})$, in which  $\dot{\varphi}_0\simeq\sqrt{C_1}$, 
see eq.(\ref{eq18}).

\section{Scale invariant coupling of $\sigma$ and $\varphi$ fields:
  decay rate of the $\sigma$ particles  to fermions}
In this section, 
we want to analyze  now a coupling of the field  $\sigma$
to a fermionic spin $1/2$ field $\Psi$. We will consider possible couplings while respecting scale invariance.
Let us consider first the  $\Psi$ kinetic term coupled to the measure $\Phi_{1}$
\begin{equation}
S= \int \Phi_{1} L_{kin},
\end{equation}
where $ L_{kin}$ is given by,
\begin{equation}
 L_{kin}= 
 \frac{i}{2}\bar{\Psi}(\gamma^a e_{a}^{\mu}
 {\overrightarrow{\bigtriangledown}}_{\mu}\Psi - \bar{\Psi}{\overleftarrow{\bigtriangledown}}_
 {\mu}\gamma^a e_{a}^{\mu} \Psi),
\end{equation}
where 
\begin{equation}
{\overrightarrow{\bigtriangledown}}_{\mu}\Psi = \partial_{\mu}\Psi + \frac{1}{2}\omega_{\mu}^{ab}\sigma_{ab}\Psi,
\end{equation}
and 
\begin{equation}
\bar{\Psi}{\overleftarrow{\bigtriangledown}}_{\mu} = \partial_{\mu}\bar{\Psi} - \bar{\Psi}\frac{1}{2}
\omega_{\mu}^{ab}\sigma_{ab}.
\end{equation}

The $\gamma^a$ matrices are metric independent (m.i.)
while $\bar{\Psi}= \Psi ^{\dagger} \gamma^0$ is as well m.i.
Since under a scale transformation we have
$\Phi_{1} \rightarrow e^\theta \Phi_{1}$, then $S_{kin}$
is invariant under
\begin{equation}
\omega_{\mu}^{ab} \rightarrow \omega_{\mu}^{ab},
\end{equation}
\begin{equation}
\Psi \rightarrow e^{-\frac{\theta}{4}}\Psi,
\end{equation}
\begin{equation}
\bar{\Psi} \rightarrow e^{-\frac{\theta}{4}}\bar{\Psi},
\end{equation}
and 
\begin{equation}
g_{\mu \nu} \rightarrow e^{\theta}g_{\mu \nu},
\end{equation}
which is equivalent also to 
$e_{a}^{\mu} \rightarrow e^{-\frac{\theta}{2}}e_{a}^{\mu} $ and  
$e_{\mu}^{a} \rightarrow e^{\frac{\theta}{2}}e_{\mu}^{a} $.
Thus, the bilinear quantity $\bar{\Psi}\Psi$ transforms as 
 $\bar{\Psi}\Psi \rightarrow e^{-\frac{\theta}{2}}\bar{\Psi}\Psi $.

So, since $\sigma$ is invariant under scale transformations, we see that a coupling to the measure 
$\Phi_{1}$ must also require a factor $e^{\alpha\varphi}$
\begin{equation}
\sigma \Phi_{1}e^{\frac{\alpha\varphi}{2}}\bar{\Psi}\Psi.
\end{equation}
Like wise, the coupling to the measure $\Phi_{2}$ (or $\sqrt{-g}$ which transforms the same way and 
which is proportional to $\Phi_{2}$), must contain a factor of $e^{\frac{3\alpha\varphi}{2}}$
leading to an invariant term
\begin{equation}
\sigma \Phi_{2}e^{3\frac{\alpha\varphi}{2}}\bar{\Psi}\Psi.
\end{equation}
Thus, the "scale invariant Yukawa type interaction" between the field $\sigma$ and the fermions must 
include $\varphi$
in the following way,
\begin{equation}
\int \sigma( g_1  \Phi_{1}e^{\frac{\alpha\varphi}{2}}\bar{\Psi}\Psi
+g_2  \Phi_{2}e^{3\frac{\alpha\varphi}{2}}\bar{\Psi}\Psi)d^4x.
\end{equation}

To properly use this interaction, we must transform to the Einstein Frame, use the Einstein Frame metric 
$\bar{g}_{\mu \nu}$ and the Einstein Frame fermion field 
$\Psi_{e.f}$. For the case $\epsilon = b = 0$, we have that
$\bar{g}_{\mu \nu} = \chi_{1}g_{\mu \nu}$, or equivalently, $\bar{e}_{\mu}^a = \chi_{1}^{\frac{1}{2}}
e_{\mu}^a$,
and $\bar{e}^a_{\mu} = \chi_{1}^{-\frac{1}{2}}{e}^a_{\mu}$. Additionally, 
the Einstein Frame fermion field satisfies the normal Dirac equation in the curved space 
$\bar{g}_{\mu \nu}$ and must be defined as
\begin{equation}
\Psi_{e.f.} = \chi_{1}^{-\frac{1}{4}} \Psi,
\end{equation}
and one can also check that $\Psi_{e.f.}$ is scale invariant.

We now will look at the interaction terms after transformation to Einstein Frame, in different phases of the theory.
For this we look $\Psi$ field as a test field which is produced in a $\sigma$ and $\varphi$ 
background. That means that we will not consider the effects of the $\Psi$ field in the 
equation for $\chi_{1}$. In this way, we can mention that
there are two interesting cases:

1. Let us consider the limit 
$\varphi \rightarrow -\infty$, which corresponds to the inflationary period and  in this case, the constants 
$M_1$ and $M_2$ can be ignored. Therefore, we can obtain that the quantity 
$\chi_{1}= \frac{2\chi_{2}f_2}{f_1}e^{\alpha \varphi}$, under such conditions, we can look at the $g_2$ 
coupling:
$g_2 \sigma \Phi_2 e^{\frac{3}{2}\alpha \varphi}\bar{\Psi}\Psi$ becomes in E.F.
$g_2 \frac{f_1}{2}  \sigma \sqrt{\bar{g}} \bar{\Psi}_{e.f.}\Psi_{e.f.}$,
which is therefore $\varphi$ independent.

Similar effect takes place for the $g_1$ coupling in the inflationary limit $\varphi \rightarrow -\infty$ 
in which the quantity 
$g_1 \sigma \Phi_1 e^{\frac{\alpha \phi}{2}}\bar{\Psi}\Psi$ transforms to    E.F. as 
$g_1 (\frac{f_1}{2 \chi_2 f_2})^{\frac{1}{2}}  \sigma \sqrt{\bar{g}} \bar{\Psi}_{e.f.}\Psi_{e.f.}$,
which is therefore again $\varphi$ independent.

2. Now let us do the same calculation in the inflationary regime in which  we study particle creation. 
In this case, the quantity $1/\chi_1$ becomes
\begin{equation}
\frac{1}{\chi_1}= 
\frac{1}{2\chi_2}\frac{V-M_1}{U+M_2},
\end{equation}
and we can neglect $M_1$ in the numerator and $U$ in the denominator,
obtaining therefore
\begin{equation}
\frac{1}{\chi_1}= 
\frac{1}{2\chi_2}\frac{V}{M_2},
\end{equation}
which implies that $\chi_1$ results
\begin{equation}
\chi_1 = 
\frac{2M_2 \chi_2}{f_1} e^{-\alpha \varphi}.
\end{equation}
Here we see that this dependence is inverse to that of the one in the inflationary 
phase (where one can ignore the constants of integration $M_1$
and $M_2$) and as a result we will get a strong
$\varphi$ dependence of the $g_1$ and $g_2$ couplings.

Let us start with the  $g_2$ coupling:
$g_2 \sigma \Phi_2 e^{\frac{3}{2}\alpha \varphi}\bar{\Psi}\Psi$ becomes in E.F.
$$
g_2\,\chi_2 \left(\frac{f_1}{2M_2\chi_2}\right)^{\frac{3}{2}} e^{3 \alpha \varphi}  
\sigma \sqrt{\bar{g}} \bar{\Psi}_{e.f.}\Psi_{e.f.}
$$
We see a very strong growth of the coupling as $\varphi$ increases in this regime.
Now for the  $g_1$ coupling:
$g_1 \sigma \Phi_1 e^{\frac{\alpha \varphi}{2}}\bar{\Psi}\Psi$ becomes in E.F.
$g_1 (\frac{f_1}{2 \chi_2 M_2})^{\frac{1}{2}}e^{\alpha \varphi}  \sigma \sqrt{\bar{g}} \bar{\Psi}_{e.f.}\Psi_{e.f.}$,
which again grows as $\varphi$ grows.

In this form, we can define that the decay rate for $\sigma$ going into two fermions becomes 
\begin{equation}
    \Gamma(\sigma\rightarrow\Psi\Psi) = \frac{g^2 m_{\sigma}}{8\pi},\label{GG}
\end{equation}
where the $g$-coupling is given by 
\begin{equation}
g= g_1 \left(\frac{f_1}{2 \chi_2 M_2}\right)^{\frac{1}{2}}e^{\alpha \varphi} +
g_2 \chi_2 \left(\frac{f_1}{2M_2\chi_2}\right)^{\frac{3}{2}} e^{3 \alpha \varphi}.\label{g2}
\end{equation}

Here we note that the decay parameter $\Gamma$ given by eq.(\ref{GG}) increases with the growth of the 
scalar field $\varphi$, (see eq.(\ref{g2})) and then the $\sigma$-particles tend to decay at large values of 
$\varphi$.

\section{Decay rates and constraints}

In this section we can study two decay rates in order to obtain different  constraints on the parameters of 
our model. In the following, we will analyze the decay rate for the specific cases  in which the coupling 
parameters $g_1=0$ and the another $g_2\neq 0$ and vice versa.   

In this context, we   consider  the special cases in which the coupling parameter $g_1=0$ and the other 
coupling parameter $g_2\neq 0$  with which the $\Gamma$-coefficient is reduced to  
\begin{equation}
    \Gamma(\sigma\rightarrow\Psi\Psi) =\Gamma_2(\sigma\rightarrow\Psi\Psi)= c_2 g_2^2 
    e^{5\alpha \varphi},\label{G1}
\end{equation}
where the constant $c_2$ is defined as $ c_2 = \frac{\mu\beta \chi_2^2}{8\pi} 
\left(\frac{f_1}{2M_2\chi_2}\right)^{3}$. Here, we have tagged the decay rate in this situation as 
$\Gamma_2$.

For the other instance  in which the coupling parameter  $g_2=0$, we have that the decay rate results 
\begin{equation}
    \Gamma(\sigma\rightarrow\Psi\Psi) =\Gamma_1 (\sigma\rightarrow\Psi\Psi)= c_1 g_1^2 e^{\alpha \varphi},
\end{equation}
in which $ c_1 =\frac{\mu\beta }{8\pi}\left(\frac{f_1}{2 \chi_2 M_2}\right)$.

 Now, as we discussed earlier we have considered   that both densities become equivalent and this occurs  
 at the equilibrium time given by  eq.(\ref{eq1}). At least during this time,  we can consider that the inflaton 
 field $\varphi$ spends most of the time previous to the equilibrium time $t_{eq}$ in which the inflaton takes 
 the value $\varphi_{eq}$. In fact, the backreaction is unimportant for times shorter than $t_{eq}$ and then  
 we can assume that the decay rate at that   time  limit $t_{eq}$ denotes by 
 $\Gamma(\varphi=\varphi_{eq})=\Gamma_{eq}$ satisfies the condition 
 in which the particles $\sigma$  will decays to fermions $\Psi$.    

At the equilibrium time, we find that the decay rate 
for the special case in which $g_1=0$ from eqs.(\ref{P1}) and (\ref{G1}) 
can be written as 
\begin{equation}
\Gamma_2(\sigma\rightarrow\Psi\Psi)\Big |_{\varphi=\varphi_{eq}}\simeq c_2g_2^2\,
\left(\frac{3\alpha_1\,t_0^2}{2}\right)^5.\label{Gg1}
\end{equation}
Analogously, we obtain  that the decay rate $\Gamma_1$ at the equilibrium time for the special case 
$g_2=0$ results
\begin{equation}
\Gamma_1(\sigma\rightarrow\Psi\Psi)\Big |_{\varphi=\varphi_{eq}}\simeq c_1g_1^2\,
\left(\frac{3\alpha_1\,t_0^2}{2}\right).\label{Gg2}
\end{equation}

Additionally,  we will assume that during the kinetic stage the Hubble factor decreases so that its value 
is similar to the decay rate $\Gamma$. Thus,   we can consider that the scalar field $\sigma$ decayed under 
the condition $H(t_{dec})=\frac{1}{3t_{dec}}\simeq\Gamma$, where  
$t_{dec}$ corresponds to the time when the scalar field $\sigma$ decayed and as the field $\varphi$ spends 
most of the time previous to the equilibrium time $t_{eq}$ we note that  this time satisfies  the condition 
$t_{dec}<t_{eq}$.

In particular for the case in which $g_1=0$ and considering the decay rate $\Gamma_2$ given by 
eq.(\ref{Gg1}), we find that the time 
 when the scalar field $\sigma$ decayed $t_{dec}$ results
$t_{dec}\simeq\frac{1}{3\,c_2g_2^2}\left(\frac{2}{3\alpha_1\,t_0^2}\right)^5$.
Similarly, for the situation in which $g_2=0$, we obtain that the time 
$t_{dec}\simeq\frac{1}{c_1g_1^2}\left(\frac{2}{9\alpha_1\,t_0^2}\right)$, when we considered the decay 
rate $\Gamma_1$. Thus, under the condition $t_{dec}<t_{eq}$ we find that for the case $g_1=0$ we have
\begin{equation}
g_2^2>\frac{\sqrt{3}}{6\,c_2}\,\left(\frac{2}{3\alpha_1\,t_0^2}\right)^5 |\dot{\varphi_0}|,\label{qq3}
\end{equation}
and for the case in which $g_2=0$, we obtain that the lower limit for the coupling $g_1$ becomes
\begin{equation}
g_1^2>\frac{\sqrt{3}|\dot{\varphi_0}|}{9\,c_1\,\alpha_1\,t_0^2},\label{qq4}
\end{equation}
here we have considered  eq.(\ref{t1}) for the equilibrium time.
Note that in both cases the lower bounds for the coupling  parameters are proportional to the velocity at the 
end of inflation, since for large $\alpha$ we have $\dot{\varphi_0}\simeq\dot{\varphi}_{end} $.

On the other hand, in order to obtain the temperature at the equilibrium time 
$T(t=t_{eq}^*)=T_{eq}(t_{eq}^*)$, we can consider  
that previous to the equilibrium time, the scalar field $\sigma$ has totally decayed. This situation occurs   
when the densities satisfy the condition
$\rho(t_{eq}^*)\sim \rho_\sigma(t_{eq}^*)$. Here we have used the notation $t_{eq}^*$  for the  time  
when the scalar field $\sigma$ has completely decayed and thus differentiate it from the equilibrium time 
$t_{eq}$ since this time is different  according  on whether  $\sigma$ field decays or not.

 As the energy density of the background $\rho(a)$ decays during the kinetic epoch   as 
 $\rho\propto a^{-6}$ and the energy density $\rho_\sigma(a)$ as radiation i.e.,  
 $\rho_\sigma\propto a^{-4}$, we have 
\begin{equation}
\rho(t_{eq}^*)=\rho(t_{dec})\left(\frac{a(t_{dec})}{a(t_{eq}^*)}\right)^6,\,\,\,\,\mbox{and}
\,\,\,\,\rho_\sigma(t_{eq}^*)=\rho_\sigma(t_{dec})
\left(\frac{a(t_{dec})}{a(t_{eq}^*)}\right)^4,
\end{equation}
with which from condition $\rho(t_{eq}^*)\sim \rho_\sigma(t_{eq}^*)$,
we find that the temperature at the equilibrium $T_{eq}\sim\rho_\sigma^{1/4}(t_{eq}^*)$ can be written as
\begin{equation}
T_{eq}\sim\rho_\sigma^{1/4}(t_{eq}^*)=\rho_\sigma^{1/4}(
t_{dec})\sqrt{\frac{\rho_\sigma(t_{dec})}{\rho(t_{dec})}}.\label{tt}
\end{equation}

On the other hand, as we have  that the scalar field $\sigma$ decayed under the condition in which 
$H(t_{dec})\simeq\Gamma$, then we assume
that the energy density of the background $\rho(t_{dec})=6H^2=6\Gamma^2$ and  for the energy density 
$\rho_\sigma(t_{dec})$ we have 
$
\rho_\sigma(t_{dec})\simeq\frac{\sqrt{3}}{3}\,\,\Gamma\,|\dot{\varphi_0}|.
$
In this way, by using Eq.(\ref{tt}) we find that  the temperature at the equilibrium can be written as 
\begin{equation}
T_{eq}\sim 10^{-1}\,|\dot{\varphi_0}|^{3/4}\,\Gamma^{-1/4}\simeq10^{-1}\,|\dot{\varphi}_{end}|^{3/4}\,
\Gamma^{-1/4} ,\label{gammaeq}
\end{equation}
where  we have used that the velocity $\dot{\varphi_0}\simeq \dot{\varphi}_{end}$ for values of $\alpha\gg 1$.

In this form, from eq.(\ref{gammaeq}) we can analyze the temperature at the equilibrium for the specific  
cases of the coupling parameters $g_1$ and $g_2$. Thus,  
in particular for the  case in which coupling parameter $g_1=0$, we find that the temperature $T_{eq}$ becomes
\begin{equation}
T_{eq}\sim10^{-2}\,\frac{|\dot{\varphi}_{end}|^{13/4}}{c_2^{1/4}\,g_2^{1/2}\,\alpha_1^{5/4}},
\end{equation}
and combining with eq.(\ref{qq3}), we obtain a lower bound for the velocity at the end of inflation 
$\dot{\varphi}_{end}$ in terms of the  temperature at the equilibrium $T_{eq}$ given by 
\begin{equation}
|\dot{\varphi}_{end}|>10^3\;T_{eq}^2.
\end{equation}
However, this  limit gives us a lower bound on the rate $\alpha^2/\chi_2^{1/2}$ given by 
\begin{equation}
\frac{\alpha^2}{\chi_2^{1/2}}>10^3\;\frac{M_2^{1/2}\,}{M_1}\,T_{eq}^2,
\end{equation}
or
\begin{equation}
\alpha^2>10^3\;\frac{M_2^{1/2}\chi_2^{1/2}\,}{M_1}\,T_{eq}^2=10^3\;\frac{1}{2\,U_{(+)}^{1/2}}\,
T_{eq}^2\simeq10^{63}\,T_{eq}^2,\label{68}
\end{equation}
where $U_{(+)}$ is defined as  $U_{(+)}=M_1^2/(4\chi_2\,M_2)$ and it  corresponds to the present vacuum 
energy density and its value is approximately  $U_{(+)}\sim 10^{-120}$(in units of $M_{Pl}^4$), see 
ref.\cite{Guendelman:2014bva}.

From eq.(\ref{68}) we can obtain different constraints on the parameter $\alpha$ depending on the 
temperature $T_{eq}$ considered, since the lower bound for $\alpha$ is given by $\alpha>10^{31}\, T_{eq}$. 

As example, by assuming that temperature at the equilibrium corresponds to the big bang 
nucleosynthesis (BBN) temperature $T_{eq}\sim T_{BBN}$ in which  $ T_{BBN}\sim 10^{-22}$ 
(in units of $M_{Pl}$), we obtain that the lower bound for the parameter $\alpha$ results $\alpha>10^{9}$.  
Now if we assume that the temperature $T_{eq}$ corresponds to the electroweak temperature 
$T_{ew}\sim 10^{-17}$,   we obtain that the lower limit for the parameter $\alpha$ results $\alpha>10^{14}$. 
Note that these constraints for the parameter $\alpha$ are consistent with considering large values of 
$\alpha$ i.e., $\alpha\gg 1$. 

Now for the special case in which the coupling term $g_2=0$, we get that the temperature at the 
equilibrium becomes
\begin{equation}
T_{eq}\sim10^{-1}\,\frac{|\dot{\varphi}_{end}|^{5/4}}{c_1^{1/4}\,g_1^{1/2}\,\alpha_1^{1/4}},
\end{equation}
and combining with eq.(\ref{qq4}), we obtain a lower limit for  $\dot{\varphi}_{end}$ as a function of the  
temperature $T_{eq}$ given by 
\begin{equation}
|\dot{\varphi}_{end}|>10^{3/2}\;T_{eq}^2.
\end{equation}

As before, this expression gives a lower bound on the ratio $\alpha^2/\chi_2^{1/2}$ results
\begin{equation}
\frac{\alpha^2}{\chi_2^{1/2}}>10^{3/2}\;\frac{M_2^{1/2}\,}{M_1}\,T_{eq}^2,
\end{equation}
and then we have
\begin{equation}
\alpha>10^{3/4}\;\frac{1}{\sqrt{2}\;U_{(+)}^{1/4}}\,T_{eq}.
\end{equation}
As in the previous case, 
by considering that the temperature $T_{eq}\sim T_{BBN}$, we obtain that the parameter $\alpha>10^{8}$ 
and for the case in which the temperature $T_{eq}$ corresponds to the $T_{ew}$, we find that the constraint 
for  $\alpha>10^{13}$. Again, we note that these results are consistent with assuming values of  
$\alpha\gg 1$.

 On the other hand, we will obtain other  constraints on the parameters of our model, by considering at 
 least another conditions during the decay of the $\sigma$ particles.

In fact, we can consider the condition  
in which the  time when the field-$\sigma$ decayed $t_{dec}$ is such that   $t_{dec}>t_{0}$,  where the 
time  $t_{0}\simeq H(t_0)^{-1}\sim H(t_{end})^{-1}=H_{end}^{-1}$. As 
 at the end of inflationary epoch the Hubble parameter $H_{end}=(V(\varphi_{end})/6)^{1/2}=
 (V_{end}/6)^{1/2}$, in which  the effective potential at the end of inflation is 
$V_{end}=f_1\,e^{-2\alpha\varphi_{end}}/(2\sqrt{\chi_2M_2})$, with which  we find that the time $t_0$ is given by 
$t_0\simeq (6/V_{end})^{1/2}=2e^{\alpha\varphi_{end}}\,(3\sqrt{\chi_2M_2}/f_1)^{1/2}$.

In this way, considering the condition in which $t_{dec}>t_0$,
we find  an upper bound for the coupling parameter $g_2$ associated to the decay rate $\Gamma_2$
for the case in which the coupling parameter $g_1=0$ given by  
\begin{equation}
g_2<8\times 10^{3}\left[\frac{M_1}{\chi_2^{1/3}f_1\,U_{(+)}^{27/12}\,\alpha^{17/3}\, \mu}\right]^{3}.\label{m1a}
\end{equation}
Here, we have used that the  time when the field-$\sigma$ decayed for the case in which the parameter  
$g_1=0$ becomes $t_{dec}\simeq\frac{1}{3\,c_2g_2^2}\left(\frac{2}{3\alpha_1\,t_0^2}\right)^5$. 

In order to evaluate an upper limit for the parameter $g_2$, we consider the special case in which $\alpha=
10^{10}$ results $g_2<10^{132}/\mu^{3}$. For the particular case in which $\alpha=10^{15}$ the upper 
bound for $g_2$ corresponds to $g_2<10^{45}/\mu^{3}$. Note that by increasing the value of the parameter 
$\alpha$ decreases the upper bound for the coupling parameter $g_2$.
Here, as before we have considered the values $M_1=4\times 10^{-60}$ (in units of $M_{Pl}^4$), 
$U_{(+)}=10^{-120}$ (in units of $M_{Pl}^4$), $\chi_2=10^{-3}$ and $f_1=2\times 10^{-8}$, respectively\cite{Guendelman:2014bva}.

Now, for the special  case in which  $g_2=0$  and considering that the  time $t_{dec}$ is given by 
$t_{dec}\simeq\frac{1}{c_1g_1^2}\left(\frac{2}{9\alpha_1\,t_0^2}\right)$, we find that the upper bound on 
the coupling parameter $g_1$ associated to $\Gamma_1$ becomes
\begin{equation}
g_1<15\left[\frac{M_1^2}{2^{1/2}\,f_1\,U_{(+)}\,\alpha^2\,\mu}\right].\label{m1c}
\end{equation}
In particular, for the specific value $\alpha=10^{10}$ we obtain that the upper bound for  $g_1$ and it  
 to $g_1<2\times10^{-10}/\mu$ and for the value $\alpha=10^{15}$ we get the bound $g_1<10^{-20}/\mu$. 
 Again we have used the values of ref.\cite{Guendelman:2014bva}, for $M_1$, $f_1$ and $U_{(+)}$.

In this context, we will obtain a range for the coupling parameters $g_1$ and $g_2$ associated to 
the decay rates $\Gamma_1$ and $\Gamma_2$, by using  the condition 
in which the $\sigma$ field decays (at the time $t_{dec}$) before reaching equilibrium (at the time $t_{eq}$)  
wherewith $t_{dec}<t_{eq}$ and from the time condition  when the field-$\sigma$ decayed at the time $t_{dec}$ 
is greater than the time $t_0\sim
H_{end}^{-1}\sim V_{end}^{-1/2}$ i.e., $t_{dec}>t_0$. 

In this form, unifying  both time conditions, we find that the range for the parameter $g_2$ associated to 
decay parameter $\Gamma_2$
for the specific  case in which the  parameter $g_1=0$ is given by 
\begin{equation}
8\times 10^{3}\left[\frac{M_1}{\chi_2^{1/3}f_1\,U_{(+)}^{27/12}\,\alpha^{17/3}\, \mu}\right]^{3}>
g_2>\frac{3^{1/4}}{\sqrt{6\,c_2}}\,\left(\frac{2}{3\alpha_1\,t_0^2}\right)^{5/2} |\dot{\varphi}_{end}|^{1/2}
.\label{m1b}
\end{equation}
Here we have used eqs.(\ref{qq3}) and (\ref{m1a}) together with the fact that $\dot{\varphi}_{0}
\simeq\dot{\varphi}_{end}$ for large $\alpha$.

In order to find a numerical range for the parameter $g_2$, we consider the special case in which $\alpha=
10^{10}$ results $(10^7/\mu^3)(\ln[10^{-26}/\mu])^{15/2}<g_2<10^{132}/\mu^{3}$. 
For the particular case in which $\alpha=10^{15}$ the range for the coupling 
 parameter $g_2$ corresponds to $(10^{-75}/\mu^3)(\ln[10^{-11}/\mu])^{15/2}<g_2<10^{45}/\mu^{3}$. 
We note that the range for the coupling parameter $g_2$ is very large. 
Here, as before we have considered that the time $t_0=(6/V_{end})^{1/2}$ together with the values $M_1=4\times 10^{-60}$ (in units of $M_{Pl}^4$), 
$U_{(+)}=10^{-120}$ (in units of $M_{Pl}^4$), $\chi_2=10^{-3}$ and $f_1=2\times 10^{-8}$, 
respectively\cite{Guendelman:2014bva}.

Analogously, for the specific  case in which the parameter $g_2=0$,  we find  that the range for the 
coupling parameter $g_1$ can be written as 
\begin{equation}
15\left[\frac{M_1^2}{2^{1/2}\,f_1\,U_{(+)}\,\alpha^2\,\mu}\right]>g_1>\frac{3^{1/4}\,|\dot{\varphi}_
{end}|^{1/2}}{3\,t_0\,\sqrt{c_1\,\alpha_1}}.\label{m1d}
\end{equation}
Here we have considered that $\dot{\varphi}_{0}\simeq\dot{\varphi}_{end}$ together with 
limits given by  eqs.(\ref{qq4}) and (\ref{m1c}), respectively.

As before, in order to obtain a range for the parameter $g_1$, we assume  the special case in which 
the parameter $\alpha=
10^{10}$ results $(10^{-12}/\mu)(\ln[10^{-26}/\mu])^{3/2}<g_1<2\times10^{-10}/\mu$, 
where the quantity  $(\ln[10^{-26}/\mu])^{3/2}<10^2$ or $\mu>4\times10^{-36}$ in order to satisfy 
the range for the parameter $g_1$. 
For the particular case in which $\alpha=10^{15}$ the range for the coupling 
 parameter $g_1$ corresponds to $(10^{-24}/\mu)(\ln[10^{-11}/\mu])^{3/2}<g_1<10^{-20}/\mu$, with 
 $\mu>10^{-213}\simeq 0$. We note that the range for the parameter $g_1$ is very narrow 
 in relation to $g_2$.
Here, as before we have used that the time $t_0=(6/V_{end})^{1/2}$ together with the values $M_1=4\times 10^{-60}$ (in units of $M_{Pl}^4$), 
$U_{(+)}=10^{-120}$ (in units of $M_{Pl}^4$), $\chi_2=10^{-3}$ and $f_1=2\times 10^{-8}$, 
respectively\cite{Guendelman:2014bva}.


\section{Conclusion}
In this paper we have analyzed in detail the instant preheating mechanism  in a scale invariant 
two measures theory. In this frame  we have studied the instant preheating for a NO model where the 
potential associated to  inflaton field does not have a minimum. Moreover, we have assumed that this 
preheating mechanism is applied to     
 an  effective potential that presents an interaction between the inflaton field $\varphi$ and other scalar 
 field $\sigma$
given by eq.(\ref{pot1}). In our analysis, we have noted that the instant preheating and in particular the  
  particles production $\sigma$ strongly depends on the interaction between the the fields $\varphi$ and 
$\sigma$. 

From the energy density of the produced particles of the field-$\sigma$, we have obtained  two limit 
decays  that depend on the effective mass $m_\sigma$ in relation to the physical momentum.  
In the first situation, we have found that  the energy density of produced particles-$\sigma$ 
decays as radiation, in a process that we called instant radiation and it occurs when the  effective mass
satisfied the condition $m_\sigma\ll k/a$. In this stage we have observed that the backreaction of 
produced particles-$\sigma$ on the equation of motion associated to the inflaton field $\varphi$  
disappears naturally product of the  evolution of the inflaton and exponential decay of the backreaction 
term.

For the  situation in which the effective mass of the field $\sigma$ satisfies the reverse situation in which 
$m_\sigma\gg k/a$, we have analyzed the possibility  that   the energy density of  produced particles 
$\sigma$ is of the same order as the energy density of the background defining an equilibrium time. 

Further, we have studied the decay rate in the framework of the scale invariant coupling of the scalar fields 
$\varphi$ and  $\sigma$ and as this  last field decays 
to fermions. Here, after performing transition to the physical Einstein frame we have considered a 
Yukawa interaction and then we have found an expression for  the decay rate
from our  scalar field  going into 
two fermions, see eq.(\ref{GG}).
From these results we have analyzed two decay rates separately assuming the values of the 
coupling parameters associated to the decay parameters. In this analysis, 
we have found different constraints 
on the coupling parameters of the decay $\Gamma$, considering the imposed conditions from the time 
when the scalar field decayed, the equilibrium time and the initial time of the kinetic epoch. 

Additionally, we have determined the temperature at the equilibrium for the different cases of the coupling 
parameters of $\Gamma$ and as example we have compared our results with the nucleosynthesis and 
electroweak temperatures, respectively. 

Finally in this article, we have not addressed the  process of the particles production 
considering 
 other
reheating mechanisms such as gravitational particle production from  massless or heavy particles 
\cite{Fo,F1}. In this sense, we hope to return 
to this point in the near future.

\begin{acknowledgments}
R.H. was supported by
Proyecto VRIEA-PUCV N$_{0}$ 039.449/2020.

\end{acknowledgments}


\end{document}